\documentclass{moriond}

\def\be{\begin{equation}}
\def\ee{\end{equation}}
\def\bea{\begin{eqnarray}}
\def\eea{\end{eqnarray}}

\newcommand{\Photo}{\includegraphics[height=35mm]{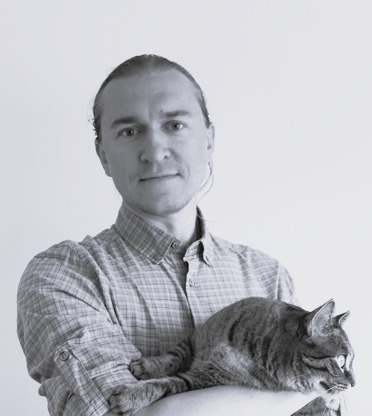}}

\begin{document}
\vspace*{4cm}
\title{Angular structure of Drell-Yan reaction in the TMD factorization approach}

\author{ A.Vladimirov}

\address{Departamento de F\'isica Te\'orica \& IPARCOS, \\ Universidad Complutense de Madrid, E-28040 Madrid, Spain}

\maketitle\abstracts{
We study angular distributions in the Drell-Yan process using an extended transverse-momentum dependent (TMD) factorization framework that includes kinematic power corrections. This approach allows the description of observables previously considered power-suppressed. The results show good agreement with experimental data and provide the first LHC-based indication of the Boer-Mulders function, highlighting the value of power corrections in TMD phenomenology.}


The production of the Drell-Yan lepton pair by the vector boson is one of the most studied cases in QCD. It serves as a baseline test for theoretical approaches and has been measured to astonishing precision in integrated and differential forms. Meanwhile the angular distributions of the Drell-Yan lepton pair are less studied. There are several reasons for it. One of them is the difficulty in measuring angular distributions in experiments.. Another is involvement of the theoretical description. Nonetheless, these angular distributions provide a wealth of information about different components of QCD, and a fine test of theoretical studies.

In this work, we considered aforementioned angular distributions in the transverse momentum dependent (TMD) factorization theorem. This approach allows to describe the $q_T$-differential cross-section in the regime $q_T\ll Q$ (with $Q$ and $q_T$ being the invariant mass and transverse momentum of the vector boson). The angle-integrated part of the cross-section is the standard study case for TMD factorization and the best source of information about unpolarized TMD parton distributions functions (TMDPDFs). The latest phenomenological studies of this part are made with the third and fourth orders of perturbative accuracy reaching the N$^4$LL level \cite{Moos:2023yfa} \cite{Moos:2025sal}. At the same time, the angular distributions remain largely unexplored, because the major part of the angular structures are power suppressed in the TMD factorization approach. Therefore, in order to describe them one should extend the TMD factorization.

\begin{figure}
\begin{minipage}{0.5\linewidth}
\centerline{\includegraphics[width=\linewidth]{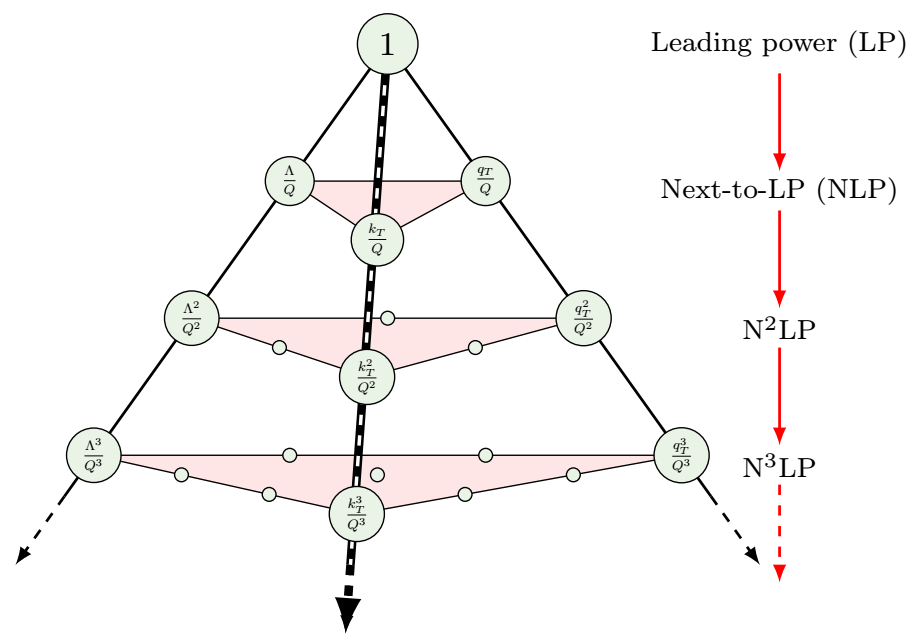}}
\end{minipage}
\begin{minipage}{0.4\linewidth}
\centerline{\includegraphics[width=\linewidth]{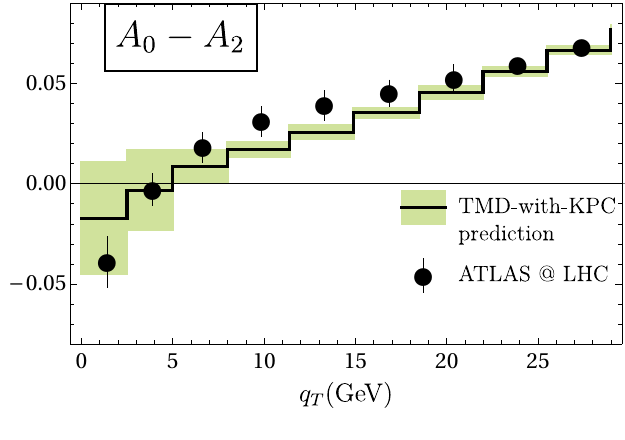}}
\end{minipage}
\caption[]{(Left) The visualization of the structure of power corrections in the TMD factorization theorem. At each power of $1/Q$-expansion (shown by rose levels) there are several types of power correction. The bold-ticked line marks KPC to the LP term, which are resummed in the TMD-with-KPC approach. (Right) The Lam-Tung angular structure measured at ATLAS \cite{ATLAS:2016rnf} compared with the prediction by TMD factorization \cite{Piloneta:2024aac}.}
\label{fig:1}
\end{figure}

The proposal of such extension appeared recently in ref.\cite{Vladimirov:2023aot}. This approach is based on the systematization of the origin of the power corrections in the TMD factorization approach. Due to the presence of several scales the power suppressed terms can be assigned to three different classes. The genuine power correction are the corrections which produce unique higher-twist operators at given power. The $q_T/Q$ and $k_T/Q$ power corrections are the corrections formed from the operator of lower-twist whose dimension is increased either by a derivative ($k_T/Q$ corrections), either by a power-suppressed part of the coefficient function ($q_T/Q$ corrections). At each order of $1/Q$-expansion all three categories are present, and beyond the next-to-leading power (NLP) approximation there are also mixed-kind of power corrections. The schematic representation of this structure is shown in fig.\ref{fig:1}(left).

Among various power corrections the $k_T/Q$ correction (also known as kinematic power corrections (KPC)) are most important. These power corrections are responsible for the restoration of frame- and gauge-invariance, which are broken at any fixed-power order. Each unique genuine and $q_T/Q$ term of power expansion has its own series of KPCs. The non-perturbative and perturbative content of a series of KPCs is identical to those of the first term. In some sense KPCs are an irreducible part of the factorization theorem and should not be dropped. In ref.\cite{Vladimirov:2023aot} the series of KPC for the LP term were studied and all these properties were confirmed by explicit computation. Furthermore, KPC for LP term were summed at all powers, and an extended version of the TMD factorization theorem (for simplicity called TMD-with-KPC factorization theorem) was presented.

In comparison to the pure LP term the TMD-with-KPC approach has the following practical features. It has the same validity range, i.e. it requires $q_T\ll Q$, and KPC are non-zero at $q_T=0$. At asymptotically large $Q$ it exactly reproduces the LP approximation. The deviations from it are $\sim 1\%$ at $Q\sim M_Z$ and grows to $\sim 5-10$\% at $Q\sim 10$GeV. Therefore, one can still use the TMDPDFs extracted by pure LP approximation (for this study we have used the ART23 TMDPDFs by ref.\cite{Moos:2023yfa}), although they should be updated for the precise phenomenology. The more important feature is that TMD-with-KPC approach gives predictions for structure functions which are zero at LP. Such as the most part of Drell-Yan angular distributions. This makes the angular distributions a perfect testing ground for the first phenomenological application of the new theory.

The standard decomposition of the Drell-Yan cross-section into angular distributions is given by
\begin{eqnarray}\label{def:angular}
\frac{d\sigma}{d^4qd\Omega}&=&\frac{3}{16\pi} \frac{d\sigma}{d^4q}\Big[
(1+\cos^2\theta)+\frac{1-3\cos^2\theta}{2}A_0+\sin2\theta \cos \phi A_1
+\frac{\sin^2\theta \cos2\phi}{2}A_2
\\\nonumber && 
+\sin\theta \cos\phi A_3+\cos\theta A_4
+\sin^2\theta \sin 2\phi A_5+\sin2\theta \sin\phi A_6+\sin\theta\sin\phi A_7\Big]
\end{eqnarray}
where $A_n$ are the desired angular distributions. The distributions $A_{4,5,6,7}$ are P-violating and thus present only for the Z-boson production. In ref.\cite{Piloneta:2024aac} the expressions for $A$'s in the TMD-with-KPC approach were found. They are expressed via the twist-two TMDPDFs: the unpolarized TMDPDF ($f_1$) and the Boer-Mulders TMDPDF ($h_1^\perp$). All distributions (except $A_1$ discussed below) are very nicely described by KPC approximation. Each of the discussed distributions demonstrates a unique feature, and thus each of them deserve a special discussion (see ref.\cite{Piloneta:2024aac}). Here, we briefly summarize some of them.

Arguably the most intriguing distribution is $A_2$. At LP it is described by the product of Boer-Mulders functions, while the unpolarized contribution is suppressed, i.e. schematically it has a form $A_2\sim h_1^\perp h_1^\perp/M^2+f_1 f_1/Q^2$. Both these terms are very clearly pronounced in the experimental measurement, with $h_1^\perp h_1^\perp$ being dominating $q_T<5$GeV part, and $f_1 f_1$ for the remaining part. It allows the first estimation of Boer-Mulders function using the LHC data. Despite large uncertainties, the Boer-Mulders function is found to be non-zero. To best of our knowledge it is the first twist-three effect (since the Boer-Mulders function is given by twist-three distribution $E(-x,0,x)$ in the collinear limit) at LHC. The more detailed investigation of this data is required. In principle, the Boer-Mulders term is present many angular distributions, but for the rest its contribution is shadowed by the pure unpolarized term.

\begin{figure}
\begin{minipage}{0.95\linewidth}
\centerline{\includegraphics[width=\linewidth]{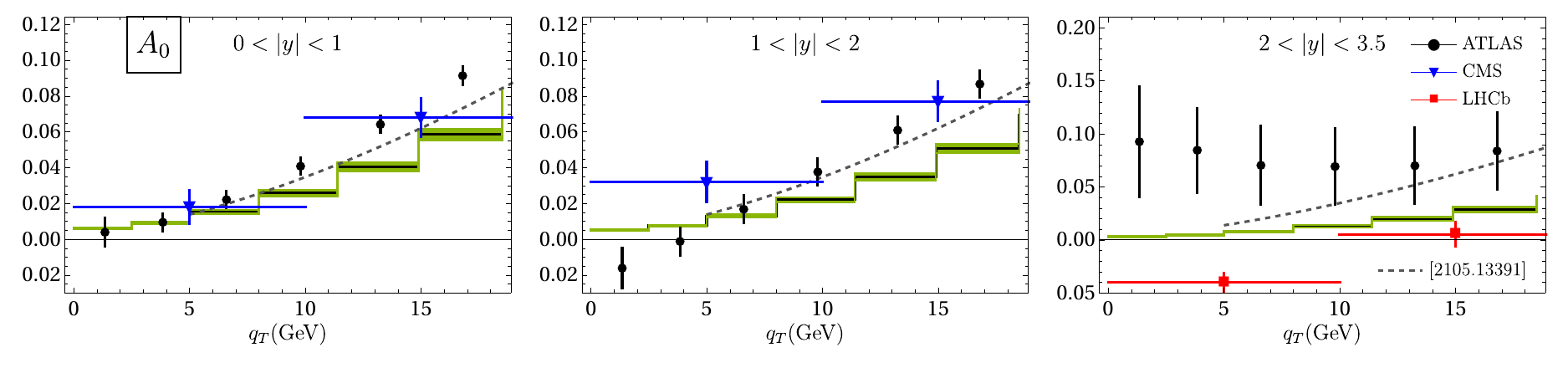}}
\end{minipage}
\caption[]{The comparison of the angular coefficient $A_0$ measured at different rapidity in refs.\cite{ATLAS:2016rnf} \cite{CMS:2015cyj} \cite{LHCb:2022tbc} and the prediction of TMD-with-KPC approach \cite{Piloneta:2024aac} (solid line with band) and pure NLP approximation \cite{Balitsky:2021fer} (dashed line).}
\label{fig:2}
\begin{minipage}{0.95\linewidth}
\centerline{\includegraphics[width=\linewidth]{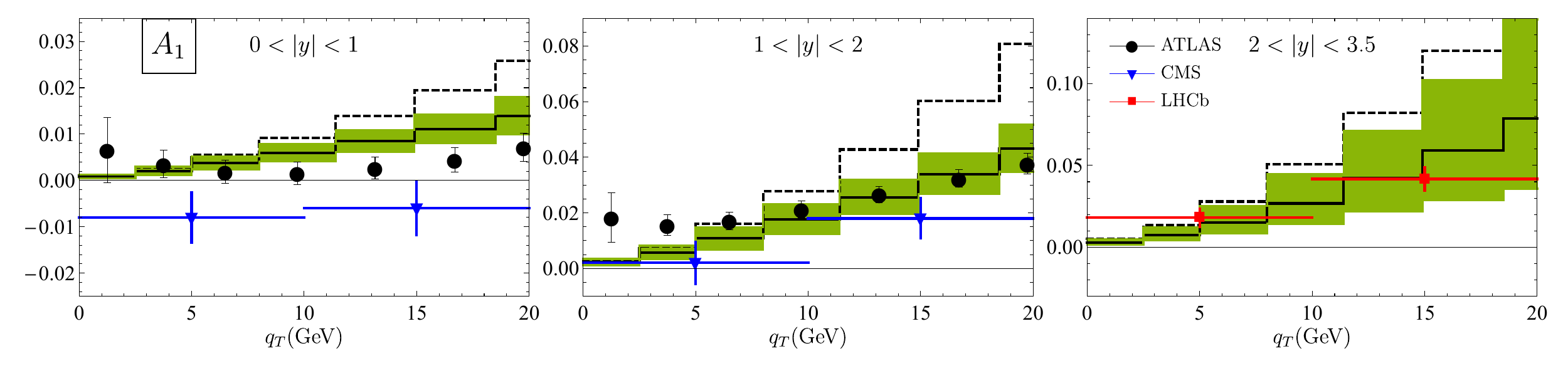}}
\end{minipage}
\caption[]{The comparison of the angular coefficient $A_1$ measured at different rapidity in refs.\cite{ATLAS:2016rnf} \cite{CMS:2015cyj} \cite{LHCb:2022tbc} and the prediction of TMD-with-KPC approach \cite{Piloneta:2024aac} (dashed line) and with addition of the $q_T/Q$ correction \cite{Arroyo-Castro:2025slx} (solid line with band).}
\label{fig:3}
\end{figure}

The distribution $A_0$ is zero at first two powers of TMD factorization approach \cite{Arroyo-Castro:2025slx}. Nonetheless the experimental measurement observes a signal at order $\sim 4-6$\% at $q_T\sim10$GeV, which is the region of TMD factorization. This signal is nicely described by KPCs, see fig.\ref{fig:2}. The same conclusion has been made in ref.\cite{Balitsky:2021fer}. The only problematic region is the large-rapidity region $2<y<3.5$, where ATLAS\cite{ATLAS:2016rnf} found an unrealistically large $A_0$ in contradiction to LHCb measurement\cite{LHCb:2022tbc}. Our prediction lies in between these two measurement (right panel in fig.\ref{fig:2}).

The finest test of KPC is given by the so-called Lam-Tung relation $A_0-A_2$. This observable is found to be significant $\sim 5$\% for $q_T<20$GeV, and very difficult to describe by the collinear factorization approach, where the leading one-loop contribution exactly cancels. The TMD-with-KPC factorization is in almost perfect agreement with measurement (see right panel in fig.\ref{fig:1}). The first two points are the contribution of the double-Boer-Mulders effect, while the rest is the contribution of unpolarized part (which is formally $1/Q^2$ suppressed). The prediction agrees with the data up to $q_T\sim 30$GeV, which is beyond the usual applicability range of the TMD factorization. This is because the Y-term contribution, which described the large-$q_T$ part, is suppressed by $\alpha_s^2$.

Finally, let us mention the angular coefficient $A_1$. This is the only coefficient found to be in disagreement with the prediction of KPCs. This is because this distribution (and also $A_3$ distribution) receives the correction $\sim q_T/Q$, which is significantly larger that $q_T^2/Q^2$ correction typical for other cases. The $q_T/Q$ correction has been computed in ref.~\cite{Arroyo-Castro:2025slx}, and together with KPC produce a very good description of the data, see fig.\ref{fig:3}. 

It must be noted that the $q_T/Q$ correction is not explicitly presented in the NLP of the TMD factorization theorem. To reveal it one must study the TMD distributions of higher-twist (twist-three in this case). These distributions exhibit a singularity in the limit of large transverse momentum. Instead usual decay $\sim 1/k_T$, they grow as $k_T^{n}$ (with $n=0$ for twist-three). This singularity can be isolated and parametrized in the terms of the twist-two TMD distributions. In this way, one separates the $q_T/Q$ correction from the true genuine correction. This is referred to as the leading-TMD approximation, and it offers an opportunity to extend the TMD formalism in the region of larger $q_T$.

\textit{Conclusion.} The TMD factorization approach has a huge hidden potential, which can be revealed via investigation of the power corrections. The first steps in this directions have been taken by the determination and summation of the kinematic power corrections (KPCs) to the leading power term, which lead to new, explicitly Lorentz invariant form of the TMD factorization theorem. This novel form enables the description of observables which are power suppressed, such as the dominant part of the angular coefficients for Drell-Yan lepton pair. In ref.\cite{Piloneta:2024aac} we made the first phenomenological test of TMD-with-KPC formalism and found that its predictions are in agreement with experimental measurements. In future, we plan to extend this approach and perform update the extraction of TMDPDFs.

\section*{Acknowledgments}

This project is supported by the Atracci\'on de Talento Investigador program of the Comunidad de Madrid (Spain) No. 2020-T1/TIC-20204 and Europa Excelencia EUR2023-143460, MCIN/AEI/10.13039/501100011033/, from Spanish Ministerio de Ciencias y Innovaci\'on.

\section*{References}
\bibliography{moriond}


\end{document}